\documentclass[useAMS,usenatbib]{mn2e}

\usepackage{graphicx}
\usepackage{dcolumn}
\usepackage{bm}
\usepackage{amsmath}

\title[Possible Effects of Pair Echoes on GRB Afterglow Emission]
{Possible Effects of Pair Echoes on Gamma-Ray Burst Afterglow Emission}
\author[Murase, Zhang, Takahashi, \& Nagataki]
{Kohta Murase$^{1}$\thanks{E-mail: kmurase@yukawa.kyoto-u.ac.jp},
Bing Zhang$^{2}$,
Keitaro Takahashi$^{1}$, 
and Shigehiro Nagataki$^{1}$
\\
$^{1}$Yukawa Institute for Theoretical Physics, Kyoto University, Sakyo-ku, Kyoto 606-8502, Japan\\
$^{2}$Department of Physics and Astronomy, University of Nevada at Las Vegas, Las Vegas, NV 89154, USA}
\begin{document}

\date{}

\pagerange{\pageref{firstpage}--\pageref{lastpage}} \pubyear{2008}

\maketitle

\label{firstpage}

\begin{abstract}
High-energy emission from gamma-ray bursts (GRBs) is widely expected but
had been sparsely observed until recently when the \textit{Fermi} satellite
was launched. If $>$~TeV gamma rays are produced in GRBs and can escape 
from the emission region, they are attenuated by the cosmic infrared 
background photons, leading to regeneration 
of $\sim$~GeV-TeV secondary photons via inverse-Compton scattering. 
This secondary emission can last for a longer time than the duration of GRBs, 
and it is called a pair echo. 
We investigate how this pair echo emission affects spectra and light curves of
high energy afterglows, considering not only prompt emission but also 
afterglow as the primary emission. Detection of pair echoes is possible as long as 
the intergalactic magnetic field (IGMF) in voids is weak. We find
(1) that the pair echo from the primary afterglow emission can affect 
the observed high-energy emission in the afterglow phase after the jet break, 
and (2) that the pair echo from the primary prompt emission can also be relevant, 
but only when significant energy is emitted in the TeV range, typically 
${\mathcal{E}}_{\gamma, >0.1~{\rm TeV}} > Y {(1+Y)}^{-1} \epsilon_{e} 
{\mathcal{E}}_{k}$. Even non-detections of the pair echoes
could place interesting constraints on the strength of IGMF.
The more favorable targets to detect pair echoes may be the ``naked" 
GRBs without conventional afterglow emission, although energetic naked
GRBs would be rare. If the
IGMF is weak enough, it is predicted that the GeV emission extends
to $> 30-300$ s. 
\end{abstract}


\section{Introduction}
High-energy emission from gamma-ray bursts (GRBs)
has been expected and various theoretical possibilities 
have been discussed by numerous authors 
\citep[see e.g.,][and references there in]{FP08}. 
In fact, EGRET detected several GRBs with GeV emission \citep[e.g.,][]{Hur+94}. 
Recently, the \textit{Fermi} satellite was launched and the onboard
Large Area Telescope (LAT) is widely expected to detect high-energy 
($>$ GeV) emission from a fraction of GRBs. 
In addition, other space- and ground-based gamma-ray observatories such
as AGILE, MAGIC, VERITAS and HESS also regard GRBs as one of the
main scientific targets. 
Theoretically there are the two main classes as high-energy emission 
mechanisms, i.e., leptonic and hadronic mechanisms. 
The leptonic mechanisms include synchrotron self-Compton
(SSC) emission and external inverse-Compton emission, which are 
the most discussed scenarios for both the prompt and the afterglow 
emission components. High-energy SSC emission is produced by relativistic 
electrons that radiate seed synchrotron photons \citep[e.g.,][]{SE01,ZM01,GG03}. 
In addition, there are various possibilities for external inverse-Compton 
emission. For example, prompt gamma-ray photons or the X-ray flare 
photons may act as seed photons for the
relativistic electrons accelerated during the afterglow phase in the
external shocks \citep[e.g.,][]{Bel05,Wan+06}.  
The hadronic mechanisms include synchrotron radiation 
of high-energy baryons, synchrotron radiation of the secondary leptons 
generated in photohadronic interactions, as well as the photons directly 
produced from $\pi^0$ decays.
In order to see the baryon synchrotron radiation, sufficiently 
strong magnetic fields are typically required \citep[e.g.,][]{GZ07,Mur+08a}. 
Otherwise, photohadronic components would dominate over the baryon
synchrotron component as long as the photon density is high enough. 
Hadronic gamma rays can be observed only when 
the nonthermal baryon loading is large enough \citep[e.g.,][]{MN06,AI07}. 
So far, both emission mechanisms have been widely considered 
in the standard scenario \citep[see reviews, e.g.,][]{Mes06,Zha07},
i.e., the internal shock model for the prompt emission and the 
external shock model for the afterglow emission, respectively. 

Both mechanisms can in principle produce $>1$ TeV photons, 
although 
high-energy photons may not escape from the source due to two-photon
pair production,
especially during the prompt emission phase \cite{LS01,GZ08,MI08,Gra+08}. 
Even if such super-TeV photons can escape from the source, they 
still suffer from pair creation due to the interaction with the 
cosmic infrared background (CIB) or the
cosmic microwave background (CMB). In particular, the direct 
detection of TeV photons would be difficult for 
GRBs with redshift $z > 1$. 
On the other hand, the electron-positron pairs resulting from the pair 
creation are still energetic, so that they up-scatter numerous CMB photons 
via the inverse-Compton process. Such secondary photons are able to 
reach the observer in a longer duration than the duration of primary 
emission, and a significant fraction of them may be observed with a 
time delay due to several effects such as magnetic deflection and angular 
spreading. Therefore, this emission is called "pair echo" emission, with 
a typical energy in the range of $\sim (1-100)$~GeV.
This pair echo emission is not only indirect evidence of the 
intrinsic TeV emission but also a clue to probe
the weak intergalactic magnetic field (IGMF) of 
$B_{\rm IG} < {10}^{-16}$ G \cite{Pla95}. 

The Plaga's method is hitherto the only one to probe very weak 
magnetic fields of $B_{\rm IG} < {10}^{-16}$ G. Other methods 
utilizing Faraday rotation or cosmic microwave background 
are sensitive to magnetic fields of order 
$B_{\rm IG} \sim 1 ~ {\rm nG}$ \cite{Kro94}. 
The presence of very weak IGMFs 
has been predicted by several mechanisms, such as 
inflation \citep[e.g.,][]{TW88}, reionization 
\citep[e.g.,][]{GFZ00} and density fluctuations
\citep[e.g.,][]{Tak+05,Ich+06}.
Observations of IGMFs in voids would give important 
information on the origin of the galactic magnetic 
fields \citep{Wid02}, although they may 
be contaminated by astrophysical
sources such as galactic winds 
or quasar outflows \cite{FL01}.

In this paper, we reinvestigate the observational effects
of the possible pair echo emission of GRB high-energy emission in 
the afterglow phase.
Three criteria should be satisfied to detect pair echo emission: 
(1) the object must emit $\sim$ TeV gamma rays leading to pair echoes; 
(2) the pair echo flux must be higher than the detector's flux sensitivity;
and (3) the pair echo emission component must not be masked by other emission components.
Concerning the point (1), TeV photons from GRBs can be emitted 
during both the prompt and the afterglow phases.
Here we consider both as the primary emission components for the echoes, 
by acknowledging that during the prompt phase
strong TeV gamma rays are
expected only for a small fraction of GRBs 
due to the large $\gamma \gamma$ optical depth, 
as has been studied by various authors \cite{DL02,MAN07,RMZ04,Tak+08}.
Concerning the point (2), we need to evaluate 
the pair echo flux quantitatively. This flux 
depends on the amount of the CIB photons, the IGMF strength,
and the source distance. As for the CIB, 
we use the acceptable CIB models given by Kneiske et al. (2002, 2004).
In order to take into account of the effects of the IGMF properly, 
we adopt the formulation developed by Ichiki et al. (2008), which 
enables us to calculate the time-dependent spectra better than the
previous works \cite{Dai+02,DL02,RMZ04,Wan+04,MAN07}.  
In addition, we have also taken into account up-scatterings of 
the CIB photons as well as the CMB photons.
This effect was neglected in the previous work for 
simplicity \cite{Tak+08}, but it can be also important \cite{MAN07}. 
In this work, we focus on the detectability of the \textit{Fermi} LAT, 
which is the most suitable one for our purpose, but also touch upon
the capabilities of other ground based TeV detectors such as MAGIC
and VERITAS. 
Concerning the point (3), we pay special attention to the high-energy 
afterglow emission, which is the main competitor of the 
pair echoes, and compare the its strengths with respect to the
echo components. Such a comparison was not done for previous
researchers who studied the pair echo.
At present, a detailed comparison between the pair echoes and high-energy 
afterglows is highly uncertain, as both have never been clearly 
detected. Since various predictions of high-energy emission rely on many 
model assumptions, they should be tested by observations of \textit{Fermi}, 
MAGIC, VERITAS and other detectors. Despite of these uncertainties,
we think it would be interesting and important to study effects of 
pair echoes that can affect high-energy emission, especially in the late 
phase \cite{DL02,RMZ04,MAN07,Tak+08}. 

\section{Emission Characteristics}
\subsection{GRB Primary Emission}
For a typical long-duration GRB, prompt gamma-ray emission is observed in a
duration of $\Delta T \sim (10-100)$ s. The typical isotropic energy 
is around $\mathcal{E}_{\gamma}^{\rm iso} \sim {10}^{53}~{\rm ergs}$. 
The observed specific flux spectrum is well approximated by a broken power-law, 
$F_{\gamma} \propto {(E_{\gamma}/{E}_{\gamma}^{b})}^{-\alpha+1}$ for 
$E_{\gamma} < E_{\gamma}^{b}$ and $F_{\gamma} \propto
{(E_{\gamma}/{E_{\gamma}}^{b})}^{-\beta+1}$ for $E_{\gamma}^{b} 
< E_{\gamma}$, where $E_{\gamma}^b$ is the break energy which 
is typically $\sim 300$ keV. $\alpha$ and $\beta$ are the low- and 
high-energy photon indices, respectively. In this work, 
we extrapolate this spectrum to higher energies and 
adopt $F_{\gamma} \propto 
{(E_{\gamma}/{E_{\gamma}}^{b})}^{-\beta+1}$ for 
$0.1~{\rm TeV} < E_{\gamma} < E_{\gamma}^{\rm cut}$, 
where $E_{\gamma}^{\rm{cut}}$ is the intrisic 
cutoff energy which is typically determined by 
the opacity of pair production. Whether TeV gamma rays 
can escape from the source strongly 
depends on the Lorentz factor and the emission radius.
Only when these quantities are large, do we expect 
TeV gamma rays escaping from the source, i.e., 
$E_{\gamma}^{\rm cut} > 1$ TeV.
Notice that although the SSC or possible hadronic mechanism
leads to more complicated spectra \citep[e.g.,][]{GG03,GZ07,AI07}, this simplification 
is sufficient for calculating the pair echo \citep[e.g.,][]{MAN07}. 
The pair echo is a kind of regenerated processes, which is composed 
of up-scattered CMB and CIB photons. The resulting 
pair echo spectrum sensitively depends on the intrinsic cutoff energy, 
while it is not so sensitive to source electron spectral indices of 
$p<3$ for a given $E_{\gamma}^{\rm cut}$ \cite{MAN07}. 
When the intrinsic cutoff energy is low enough, the resulting 
spectrum basically reflects the seed CMB and CIB spectra\footnote{If the spectrum of 
relativistic pairs is expressed by a power-law with an index of $s$, 
the inverse-Compton spectrum is expected as $F_{\gamma} \propto 
\varepsilon^{-\frac{s-1}{2}}$ below the peak. However, the pair 
spectrum is strongly affected by the CIB field, and is proportional 
to $(1-e^{-\tau_{\gamma \gamma} (E_{\gamma},z)})$, where 
$\tau_{\gamma \gamma} (E_{\gamma},z)$ is the optical depth of photons 
with $E_{\gamma}$ emitted at the redshift $z$. 
Since the pair echo spectrum is rather sensitive to the IGMF 
and the CIB spectrum, it is not easy to know 
a source electron spectral index $p$.}, which roughly leads to 
the spectral peak of $\sim ((1+z) E_{\gamma}^{\rm cut}/2 m_e c^2)^2 
k_B T_{CMB}^{\prime}/(1+z)$. Here, $T_{\rm CMB}^{\prime}=2.73 
(1+z)$ K is the local CMB temperature. On the other hand, 
when the intrinsic cutoff energy is high enough, high-energy 
secondary photons are re-absorbed, and the resulting spectrum 
has the cutoff due to CMB/CIB absorption. As the intrinsic 
cutoff energy is higher, the cascade effect becomes more and more 
significant, i.e., repeating the pair creation and inverse-Compton 
scattering is important. It affects the resulting spectrum, erasing 
the memory of the primary spectrum in the high energies.
Rather, the radiation energy output above TeV is 
important for the pair echo flux, and we normalize 
the primary flux through the isotropic radiation energy above 0.1 TeV,
$\mathcal{E}_{\gamma, >0.1~{\rm TeV}}$.  

The prompt emission is followed by the afterglow phase,
during which the relativistic 
ejecta is decelerated by a circumburst medium.   
A pair of external shocks (forward and reverse)
form, from which electrons (and 
possibly baryons) are accelerated and radiate afterglow photons. 
High-energy emission during this phase was predicted by 
many authors in both of the reverse and forward shock models. 
\cite[see][and references there in]{FP08}.
TeV emission in the external shocks has a smaller optical depth
for pair production, and hence, can escape the source more easily.
For the forward shock, 
the characteristic energies for the SSC emission 
are given by \citep[e.g.,][]{SE01,ZM01,GG03}
\begin{eqnarray}
E_{\rm SSC}^m &\simeq& 2.3 \times {10}^{3}~{\rm eV}~g_{-1}^4 
~{\epsilon}_{e,-1}^4~{\epsilon}_{B,-2}^{\frac{1}{2}}
~\mathcal{E}_{k,53}^{\frac{3}{4}}~n_0^{-\frac{1}{4}}
~t_4^{-\frac{9}{4}} \\
E_{\rm SSC}^c &\simeq& 2.2 \times {10}^{10}~{\rm eV}
~{\epsilon}_{B,-2}^{-\frac{7}{2}}~{\mathcal{E}}_{k,53}^{-\frac{5}{4}}
~n_0^{-\frac{9}{4}}~{(1+Y)}^{-4}~{t}_{4}^{-\frac{1}{4}},
\end{eqnarray}
where $\epsilon_B$ and $\epsilon_e$ are the fractions of the shock energy 
transferred to the downstream magnetic fields and nonthermal electrons, 
respectively. $g=g(p)$ is a numerical factor, which is expressed as 
$g(p)=(p-2)/(p-1)$ for $p>2$ and the typical value for $p \sim 2$ 
is $g \sim 0.1$. $\mathcal{E}_{k}$ is the isotropic kinetic energy of
the ejecta, $n$ is the circumburst medium density\footnote{
We focus on the uniform medium in this work.
}, and 
$Y$ is the Compton parameter.  
For $\epsilon_e > \epsilon_B$, we roughly 
have\footnote{$Y \sim 
\sqrt{\epsilon_e/\epsilon_B}$ is expected when 
only the first SSC component is important. In fact, 
the second SSC component is typically negligible 
due to the Klein-Nishina suppression in 
the optically thin synchrotron scenario.}  
$Y \sim \sqrt{\epsilon_e/\epsilon_B}$ 
\citep[e.g.,][]{SE01,ZM01}, and 
the high-energy emission spectrum is written as
$F_{\rm SSC} \propto {E_{\rm SSC}}^{1/3}$ for 
$E_{\rm SSC} < E_{\rm SSC}^{m}$, $F_{\rm SSC} \propto
{E_{\rm SSC}}^{-(p-1)/2}$ for $E_{\rm SSC}^{m} < E_{\rm SSC} < 
E_{\rm SSC}^c$ and $F_{\rm SSC} \propto
{E_{\rm SSC}}^{-p/2}$ for $E_{\rm SSC}^{c} < E_{\rm SSC} < 
E_{\rm SSC}^{\rm cut}$, where $p \sim 2-3$ is the spectral index
of the accelerated electrons. 
Here $E_{\rm SSC}^{\rm cut}$ is the cutoff energy determined 
either by the pair-creation opacity or the Klein-Nishina limit 
\citep[e.g.,][]{ZM01}. 
The energy flux at the SSC peak (for $p \sim 2$) is evaluated as
\begin{eqnarray}
E_{\rm SSC}^c~F_{\rm SSC}^{c}~&\simeq&~2.7~\times~{10}^{-8}
~{\rm GeV}~{\rm cm}^{-2}~{\rm s}^{-1} \nonumber \\
&\times&~Y {(1+Y)}^{-1} g_{-1} \epsilon_{e,-1}
\mathcal{E}_{k,53} t_{4}^{-1} D_{28}^{-2},
\end{eqnarray}
by which we can normalize the SSC spectrum. The 
above temporal behavior is typically valid
from the break time of $t_b \sim {10}^4$ s to 
the next break time of $t_{j} \sim {10}^{5}$ s during
the so-called normal decay phase of X-ray afterglow. 
Afterglow light curves of some GRBs are steepened 
after $t_j$, which is often interpreted as a jet break 
when the Lorentz factor $\Gamma$ becomes the inverse 
of the jet opening angle\footnote{Notice that
the predicted achromaticity of this jet-like break is only 
verified for a fraction of GRBs \citep{Liang08}} $1/\theta_j$
(Rhoads 1999; Sari et al. 1999).
The temporal behavior after the jet break $t_j$ is expected as
$E_{\rm SSC}^m \propto t^{-3}$, $E_{\rm SSC}^c \propto 
t^{1}$, $E_{\rm SSC }^{\rm cut} \propto t^{-1/2}$ 
and $E_{\rm SSC}^c F_{\rm SSC}^c \propto t^{-2}$.

The afterglow behavior before $t_b$ cannot be interpreted
by the standard afterglow model. 
As observed by \textit{Swift}, a good fraction of X-ray afterglow
has a shallow decay phase
lasting from $t_a \sim {10}^{3}$ s (at which the shallow 
decay emission becomes dominant in x rays) to $t_b \sim {10}^{4}$ s
\citep[see, e.g.,][]{Nou+06,OBr+06},
which has a decay slope of $\propto t^{-(0-0.8)}$.
Several models have been proposed for explaining this phase 
\citep[see, e.g.,][]{Zha+06,EG06,GDM07,Ghi+07,Pan07,UB07,Yam09}, 
and one of the mostly discussed interpretations is continuous energy 
injection into the forward shock. Here we consider 
the modified forward shock model with the energy injection of the form 
$\mathcal{E}_k \propto t^{1-q}$, where $q$ parameterizes the energy 
injection and $q=1$ corresponds to the case of no energy injection.
Such modified forward shock models are supported by the lack
of spectral evolution across $t_b$ and the compliance of
the ``closure relations" in the normal decay phase after $t_b$
\cite{LZZ07}. During this phase, the temporal behavior 
of various parameters are 
$E_{\rm SSC}^m \propto t^{-3/2-3q/4}$, $E_{\rm SSC}^c \propto 
t^{-3/2+5q/4}$, $E_{\rm SSC}^{\rm cut} \propto t^{-q/4}$ and 
$E_{\rm SSC}^c {F}_{\rm SSC}^c \propto t^{-q}$ \cite{Fan+08}.
We have calculated the high energy light curves of the SSC
emission during this phase. Similar calculations were
performed by e.g., Gou \& M\'esz\'aros (2007), Wei \& Fan 
(2007), and Fan et al. (2008). 

\subsection{Pair Echo Emission}
Pair echoes are the up-scattered CMB and CIB photons by the
electron-positron pairs produced via the attenuation of the
primary TeV photons by the CIB.
For a given primary spectrum, 
the total fluence of the pair echo emission is determined by
the $\gamma \gamma$ optical depth of the CIB, and does not 
depend on the IGMF as long as the deflection angle is much 
smaller than the jet opening angle. Primary photons with 
energy $E_{\gamma}$
are converted to pairs with Lorentz factor 
$\gamma_e \approx {10}^{6} (E_{\gamma}/1 \, {\rm{TeV}}) (1+z)$ 
in the local cosmological rest frame,
which then up-scatter CMB and CIB photons. 
CMB photons are boosted to energies
$\sim 2.82 k_B T_{\rm{CMB}}^{\prime} \gamma_e^2 /(1+z)
\approx 0.63 {(E_{\gamma}/1 \, \rm{TeV})}^{2} {(1+z)}^{2}$ GeV.
To evaluate the pair echo flux,
we must consider various time scales involved in the process,
such as the angular spreading time,
and the delay time due to magnetic deflections \citep[e.g.,][]{DL02,Dai+02,RMZ04}.  
These can be estimated as follows \cite{Tak+08,Mur+08}. 

The angular spreading time is
${\Delta t}_{\rm{ang}} \approx (1+z) (\lambda _{\rm IC}^{\prime} + \lambda
_{\gamma \gamma}^{\prime})/2{{\gamma}_{e}}^{2}c$,
where $\lambda _{\gamma \gamma}^{\prime} \approx 
{(0.26 \sigma_T n'_{\rm{CIB}})}^{-1} \approx 
20~{\rm Mpc}~{(n'_{\rm{CIB}}/ 0.1~{\rm{cm}^{-3}})}^{-1}$
is the local $\gamma\gamma$ mean free path
in terms of the local CIB photon density $n'_{\rm{CIB}}$,
and $\lambda _{\rm IC}^{\prime}=3 m_e c^2/(4 \sigma_T U'_{\rm{CMB}} \gamma_e)
\approx 690 \, {\rm kpc} {(\gamma_e/{10}^{6})}^{-1} {(1+z)}^{-4}$
is the local IC cooling length
in term of the local CMB energy density $U'_{\rm{CMB}}$.
At the energies of our interest, $\lambda' _{\gamma \gamma} \gg \lambda' _{\rm IC}$ so that
${\Delta t}_{\rm{ang}} \approx (1+z) 
\lambda _{\gamma \gamma}^{\prime}/2{{\gamma}_{e}}^{2}c
\approx 960 \, {\rm{s}} {(\gamma_e/{10}^{6})}^{-2}
{(n'_{\rm{CIB}}/ 0.1 \, {\rm{cm}^{-3}})}^{-1} (1+z)$.
For sufficiently small deflections in weak IGMFs
with the present-day amplitude $B_{\rm IG}=B_{\rm IG}^{\prime} {(1+z)}^{-2}$ and 
coherence length $\lambda_{\rm{coh}}=\lambda_{\rm{coh}}^{\prime} (1+z)$,
the magnetic deflection angle is
$\theta_B = {\rm{min}}[\lambda_{\rm{IC}}^{\prime}/r_L,
({\lambda_{\rm{IC}}^{\prime} 
{\lambda}_{\rm{coh}}^{\prime})}^{1/2}/r_L]$,
where  $r_{\rm{L}}=\gamma_e m_e c^2/e B_{\rm IG}^{\prime}$ is the
Larmor radius of the electrons or positrons\footnote{
There was a typo on the expression of $\theta_B$ in Murase et al. 2008b. 
The ``minimum" is correct rather than the ``maximum".
The calculations were performed properly.}.
The delay time due to magnetic deflection is
${\Delta t}_{B} \approx (1+z) (\lambda _{\rm IC}^{\prime}+ 
\lambda _{\gamma \gamma}^{\prime}) ({\theta}_{B}^{2}/2c)$.
For coherent magnetic fields with
$\lambda'_{\rm{coh}} > \lambda'_{\rm{IC}}$,
we have ${\Delta t}_{B} \approx {\rm{max}}
[6.1 \times {10}^{3} \, {\rm{s}} {(\gamma_e/{10}^{6})}^{-5}
{(B_{\rm{IG}}/ {10}^{-20} \, \rm{G})}^{2} {(1+z)}^{-7}, 
1.6 \times {10}^{5} \, {\rm{s}} {(\gamma_e/{10}^{6})}^{-4} 
{(n_{\rm{CIB}}^{\prime}/ 0.1 \, {\rm{cm}^{-3}})}^{-1} 
{(B_{\rm{IG}}/ {10}^{-20} \, \rm{G})}^{2} {(1+z)}^{-3}]$.
Note that the deflection angle due to successive IC scattering
$\theta_{\rm{IC}} \approx \sqrt{N} 
k_B T_{\rm{CMB}}^{\prime}/m_e c^2$ is usually very small, where $N
\approx \lambda_{\rm{IC}}^{\prime}/l_{\rm{IC}}^{\prime} \sim 1000$ is the number of
scatterings and $l_{\rm{IC}}^{\prime}$ is the IC scattering mean free path. 
We have also assumed that both $1/\gamma_e$ and 
$\theta_B$ do not exceed $\theta_j$;
otherwise a significant fraction of photons or pairs will be
deflected out of the line of sight and the echo flux is greatly diminished.

In order to calculate the pair echo flux, we adopt the formalism developed by 
Ichiki et al. (2008), which enables us to calculate the time-dependent 
spectra in a more satisfactory manner, particularly at late times,
accounting properly for the geometry of the pair echo process. 
In previous works, explicit descriptions of the 
time-dependent spectra were not possible 
without some ad hoc modifications \cite{And04,MAN07}.

\section{Effects of Pair Echoes on High-Energy Afterglow Emission}
In this section, we present our results and compare 
the pair echo emission with the afterglow emission. 
The detectability by the \textit{Fermi}/LAT detector and 
the ground-based MAGIC telescope are 
also discussed.
One main
uncertainty stems from the CIB models,
which can affect not only the pair echo fluence but also the
time scales for angular spreading and magnetic deflection
at all redshifts. 
Recent high-energy observations of TeV blazars point to 
a low-IR CIB model, close to the lower 
limit from the galaxy count data \citep[e.g.,][]{Alb+08}
(but see, e.g., Stecker \& Scully 2008). 
Hence, we here adopt the low-IR CIB model presented by 
Kneiske et al. (2002, 2004). More 
detailed discussion on the effects of the CIB is 
found in Murase et al. (2007).
As for the afterglow parameters in the forward 
shock model, we adopt $\mathcal{E}_{k}={10}^{52-53}$ 
ergs, $\epsilon_e=0.1$, $\epsilon_B=0.01$, 
$n=1~{\rm cm}^{-3}$ and $p=2.0-2.4$.
We also assume the energy injection index $q=0.5$ 
before $t_{b}={10}^{4}$ s, 
and take the jet break time as $t_j ={10}^{5}$ s.   

\subsection{Afterglow-Induced Pair Echoes vs Afterglows}  
\begin{figure}
\includegraphics[width=\linewidth]{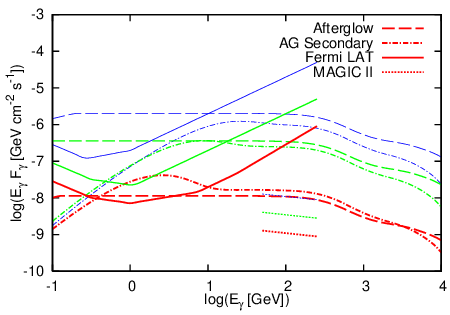}
\caption{Primary and pair echo spectra for the canonical afterglow
with $\mathcal{E}_{k}={10}^{53}$ ergs and $p=2.0$, 
plotted at $t={10}^{3.5}$ s (blue),  $t={10}^{4.5}$ s (green) 
and $t={10}^{5.5}$ s (red), for the case of 
$B_{\rm IG}=10^{-20}$ G, $\lambda_{\rm{coh}}=1$ Mpc, and $z=0.1$.
The {\em Fermi}/LAT and MAGIC II sensitivities (with the duty factor of 20 \%) 
are also overlayed (Carmona et al. 2007). Note that 
the sensitivity curves in the sky survey mode are used for the long time
observations, although the possible continuous observations by LAT may 
improve the detectability by a factor of 3-5 
(e.g., Gou \& M\'esz\'aros 2007).
}

\end{figure}
\begin{figure}
\includegraphics[width=\linewidth]{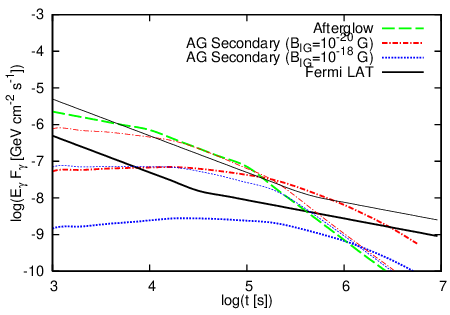}
\caption{
Primary and pair echo light curves for the canonical afterglow 
with $\mathcal{E}_{k}={10}^{53}$ ergs and $p=2.0$, 
compared with the LAT sensitivity at 1 GeV (thick) and 10 GeV (thin),
for the case of $B_{\rm IG}=10^{-20}$ G with with $\lambda_{\rm{coh}}=1$ 
Mpc and $B_{\rm IG}={10}^{-18}$ G with $\lambda_{\rm{coh}}=0.1$ kpc.
The source redshift is $z=0.1$.}
\end{figure}

\begin{figure}
\includegraphics[width=\linewidth]{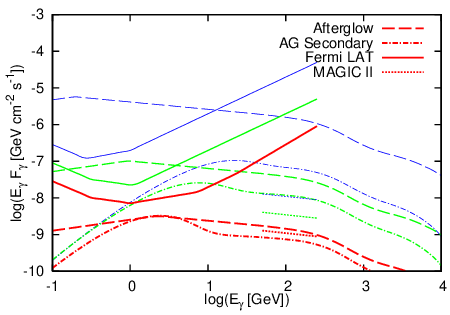}
\caption{The same as Fig. 1, but for the canonical afterglow
with $\mathcal{E}_{k}={10}^{52}$ ergs and $p=2.4$. 
}

\end{figure}
\begin{figure}
\includegraphics[width=\linewidth]{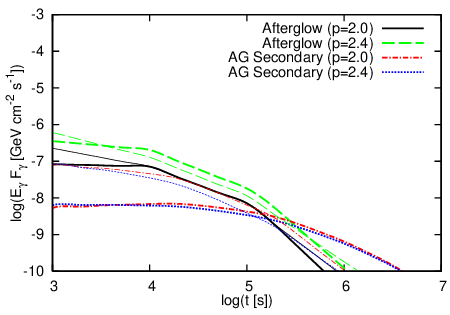}
\caption{
Primary and pair echo light curves for the canonical afterglow 
with $\mathcal{E}_{k}={10}^{52}$ ergs, for the cases of $p=2.0$ 
and $p=2.4$, respectively. 
Light curves at 1 GeV (thick) and 10 GeV (thin) are 
shown for the case of $B_{\rm IG}={10}^{-20}$ G 
with $\lambda_{\rm{coh}}=1$ Mpc. The source redshift is $z=0.1$.}
\end{figure}

In Figs. 1 and 2, we show the resulting spectra and light 
curves of the afterglow-induced pair echo and the primary 
afterglow emission. 
We can see that the echo component is 
out-shined by the afterglow component 
during the shallow and normal decay phases. This result is 
consistent with Ando (2004), who argued that observed emission
is unaffected by the pair echo. 
The situation changes dramatically after the jet break. 
The pair echo emission lasts for a long time 
because of the IGMF deflection of the pairs, 
and it can dominate the afterglow 
after the jet break by as much as an order of magnitude.
It can be observed only for nearby GRBs
with $z < 0.2$ for our afterglow parameters. 

If a GRB is very nearby and energetic, we may detect many photons
at $\sim$ GeV energies and even observe TeV photons during the 
afterglow phase.
In such a case, in principle
a non-detection of the high energy pair echo 
would allow us to obtain the lower limit on the IGMF. 
This is because if $B_{\rm{IG}}=0$
one would expect an excess of the echo flux $F_{\rm{sec}}$
over the primary flux $F_{\rm{pri}}$.
The non-detection of the echo emission can then be attributed to 
the effect of a finite IGMF, which deflects the secondary pairs 
to reduce the secondary echo flux to be 
$F_{\rm{sec}} < {\rm{max}}
(F_{\rm{pri}}, F_{\rm{lim}})$, where 
$F_{\rm{lim}}$ is the detector sensitivity \cite{Mur+08}. 
The expected lower bound with our afterglow parameters 
($\mathcal{E}_{k}={10}^{53}$ ergs and $p=2.0$) for a GRB 
with $z=0.1$ is estimated as
\begin{equation}
B_{\rm IG} \cdot {\rm min[\lambda_{\rm coh}^{1/2},\lambda_{\rm IC}^{1/2}]} > {10}^{-21}~{\rm G}~{\rm Mpc}^{1/2}. 
\end{equation}
In general the result depends on 
the source distance and the afterglow parameters 
which should be determined from 
observational properties. In any case, the expected lower bounds 
are comparable to those derived for blazar flares \cite{Mur+08}. 

Similar to the case of blazar flares, one expects that 
whether the afterglow pair echo dominates over the primary emission
depends on the high-energy afterglow spectrum. In Figs. 3 and 4, we show the 
case of $p=2.4$, corresponding to $F_{\rm SSC} \propto E_{\rm SSC}^{-1.2}$. 
Obviously, such steeper indices make it more difficult to see 
the afterglow-induced pair echo emission. This is just because 
steeper indices imply the smaller TeV flux compared to the GeV flux
as for the afterglow emission. 
Hence, the electron spectral index is one of the uncertainties 
that are closely relevant to whether the afterglow-induced 
pair echoes are detectable. 
Also, it is clear that brighter afterglows are favorable for 
detections. Since the pair echo can be dominant over 
the afterglow itself only after the jet break, we need 
to observe kind of energetic afterglows 
with $Y {(1+Y)}^{-1} \epsilon_e \mathcal{E}_k > {10}^{51.5}$ ergs 
for $z=0.1$ (see Figs. 1-4). 

\subsection{Prompt-Induced Pair Echoes vs Afterglows}
\begin{figure}
\includegraphics[width=\linewidth]{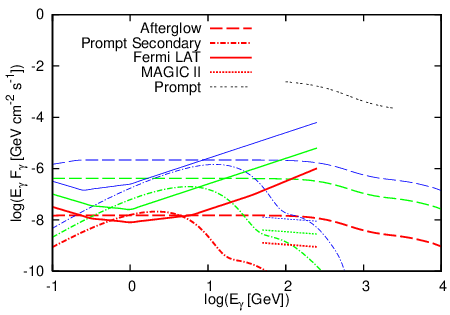}
\caption{Spectra of the afterglow and the pair echo of the prompt emission, 
plotted at $t={10}^{3.5}$ s (blue),  $t={10}^{4.5}$ s (green) 
and $t={10}^{5.5}$ s (red), for the case of 
$B_{\rm IG}=10^{-20}$ G with $\lambda_{\rm{coh}}=1$ Mpc. 
The \textit{Fermi}/LAT and MAGIC II sensitivities (with the duty factor of 20 \%) 
also plotted for comparison. 
The prompt emission spectrum at $t=0$ s is shown, with
$\mathcal{E}_{\gamma, >0.1~{\rm TeV}}={10}^{52}$ ergs assumed.
The canonical afterglow spectrum is also shown for 
the case of $\mathcal{E}_{k}={10}^{53}$ ergs and $p=2.0$, 
The source redshift is $z=0.1$.}
\end{figure}
\begin{figure}
\includegraphics[width=\linewidth]{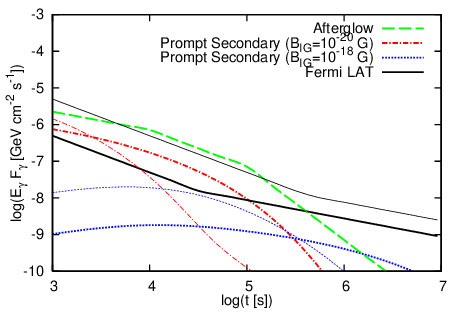}
\caption{Light curves of the afterglow and the pair echo for the prompt emission 
compared with the LAT sensitivity at 1 GeV (thick) and 10 GeV (thin),
for the case of $B_{\rm IG}=10^{-20}$ G and $B_{\rm IG}={10}^{-18}$ G 
with $\lambda_{\rm{coh}}=0.1$ kpc. Here $\mathcal{E}_{\gamma, >0.1~{\rm TeV}}
={10}^{52}$ ergs is assumed. The source redshift is $z=0.1$.}
\end{figure}
\begin{figure}
\includegraphics[width=\linewidth]{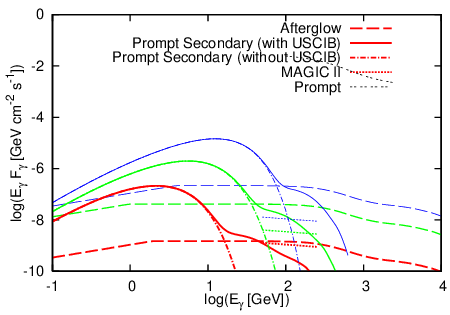}
\caption{Spectra of the afterglow and the pair echo of the prompt emission, 
plotted at $t={10}^{3.5}$ s (blue),  $t={10}^{4.5}$ s (green) 
and $t={10}^{5.5}$ s (red), for the case of 
$B_{\rm IG}=10^{-20}$ G with $\lambda_{\rm{coh}}=1$ Mpc. 
The \textit{Fermi}/LAT and MAGIC II sensitivities (with the duty factor of 20 \%) 
also plotted for comparison. 
The prompt emission spectrum at $t=0$ s is shown, with
$\mathcal{E}_{\gamma, >0.1~{\rm TeV}}={10}^{53}$ ergs assumed.
The canonical afterglow spectrum is also shown for 
the case of $\mathcal{E}_{k}={10}^{52}$ ergs and $p=2.0$, 
The source redshift is $z=0.1$. 
In order to demonstrate the effect of up-scattered CIB (USCIB) photons (solid),
curves without up-scattering of CIB photons are also shown (dot-dashed). 
}
\end{figure}
\begin{figure}
\includegraphics[width=\linewidth]{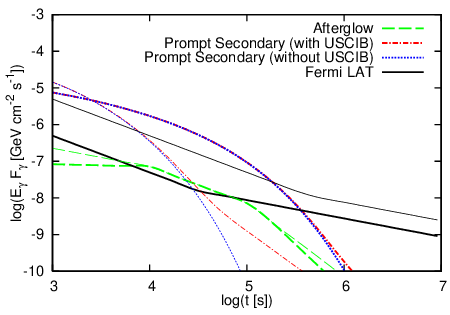}
\caption{Light curves of the afterglow and the pair echo for the prompt emission 
compared with the LAT sensitivity at 1 GeV (thick) and 10 GeV (thin),
for the case of $B_{\rm IG}=10^{-20}$ G and $B_{\rm IG}={10}^{-18}$ G 
with $\lambda_{\rm{coh}}=0.1$ kpc. Here the relevant parameters of 
the prompt emission and afterglow are the same as those used in Fig. 7.
The source redshift is $z=0.1$.}
\end{figure}

In Figs. 5 and 6, we show the resulting spectra and light 
curves of the prompt-induced pair echo. The parameters for the 
primary prompt emission are taken as the following:  
$\mathcal{E}_{\gamma, >0.1~{\rm TeV}}={10}^{52}$ ergs, $\beta=2.2$ and 
$E_{\gamma}^{\rm cut}={10}^{0.5}$ TeV. 
The duration in the local rest frame is set to $\Delta T^{\prime}=25~{\rm s}$. 
For comparison we also show the afterglow spectra/light curves. 
We notice that the prompt-induced pair echo has been discussed
by several authors before, but the comparison with the
afterglow flux was never done previously. We find that the pair 
echo is observable only when GRBs are strong TeV emitters, i.e.
$\mathcal{E}_{\gamma, >0.1~{\rm TeV}} > {10}^{52}$ ergs for 
our afterglow parameters (where $Y {(1+Y)}^{-1} \epsilon_e 
\mathcal{E}_k > {10}^{52}$ ergs). This is a strong requirement 
for the GRBs with canonical afterglows. 
For weak but non-zero IGMFs, the pair echo lasts for a longer time although 
its maximum flux is lower than the case of $B_{\rm IG}=0$. 
Then, the echo could still dominate over 
the afterglow at late times after the jet break, since 
its light curve is shallower than that of the afterglow. 

In Figs. 7 and 8, we show the more optimistic cases where 
brighter prompt emission and dimmer afterglow emission 
are assumed. In those cases, the observed behavior of 
high-energy afterglows is quite different from the 
predicted one from the afterglow theory, since 
the pair echo emission is dominant for a long time. 
A weak but non-zero IGMF with $B_{\rm IG} < {10}^{-20}$ G 
can even make the pair halo out-shine the shallow 
decay emission. In Figs. 7 and 8, 
we also show the effect of up-scattered CIB photons. 
As is easily seen, their effect is important at high 
energies above 10-100 GeV, which can be crucial for 
detections through the MAGIC and VERITAS telescopes. 
Note that this effect becomes important when 
the intrinsic cutoff energy is not so high, as pointed out
in Murase et al. 2007. Otherwise, the up-scattered CIB 
component is masked by the up-scattered CMB component. 
In fact, it is typically difficult to see the former 
for afterglow-induced pair echoes, where the pair echo 
spectrum at $t$ is composed of the up-scattered CMB photons 
produced by the primary photons emitted at different 
times from the source.  

Similar to what has been discussed in the previous subsection,
one may obtain
the lower bound on the IGMF for non-detection of the prompt-induced 
pair echo. However, the relative importance
of the prompt-induced pair echo with respect to the afterglow emission 
is complicated, which strongly depends on the ratio of the
prompt TeV emission energy and the electron energy in the afterglow
($\epsilon_e \mathcal{E}_k$).
In addition, the afterglow-induced pair echo would also contaminate 
the prompt-induced pair echo. Here, for a conservative estimate, 
let us consider the epochs of $t< t_j$. Assuming 
that TeV emission is detected, a non-detection
of the pair echo would lead to
\begin{equation}
B_{\rm IG} \cdot {\rm min[\lambda_{\rm coh}^{1/2},\lambda_{\rm IC}^{1/2}]} > {10}^{-19.5}~{\rm G}~{\rm Mpc}^{1/2}, 
\end{equation}
for our prompt and afterglow parameters used in Fig. 7. 

\section{Pair Echoes from ``Naked" GRBs}
\begin{figure}
\includegraphics[width=\linewidth]{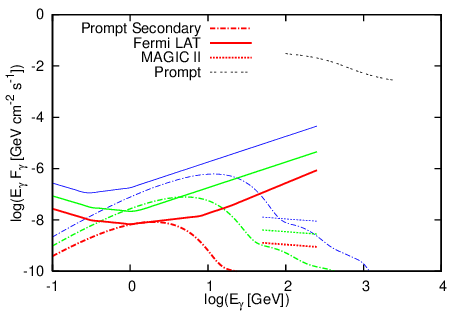}
\caption{Spectra of the pair echo of the prompt emission from a naked short GRB, 
plotted at $t={10}^{3.5}$ s (blue),  $t={10}^{4.5}$ s (green) 
and $t={10}^{5.5}$ s (red), for the case of 
$B_{\rm IG}=10^{-20}$ G with $\lambda_{\rm{coh}}=1$ Mpc. 
The \textit{Fermi}/LAT and MAGIC II sensitivities (with the duty factor of 20 \%) 
also plotted for comparison. 
The prompt emission spectrum at $t=0$ s is also shown, with
$\mathcal{E}_{\gamma, >0.1~{\rm TeV}}={10}^{51.5}$ ergs assumed.
The source redshift is $z=0.1$.}
\end{figure}
\begin{figure}
\includegraphics[width=\linewidth]{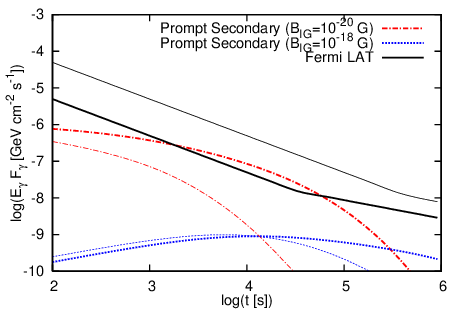}
\caption{Light curves of the pair echo for the prompt emission from a naked short 
GRB compared with the LAT sensitivity at 1 GeV (thick) and 10 GeV (thin),
for the case of $B_{\rm IG}=10^{-20}$ G and $B_{\rm IG}={10}^{-18}$ G 
with $\lambda_{\rm{coh}}=0.1$ kpc. Here $\mathcal{E}_{\gamma, >0.1~{\rm TeV}}
={10}^{51.5}$ ergs is assumed. The source redshift is $z=0.1$.}
\end{figure}

As seen in the previous subsection (see Figs. 5 and 6), afterglow emission 
may significantly mask a pair echo (for both of long and short GRBs). 
Hence, of special interest are the GRBs whose intrinsic high energy afterglow 
emission is weak and whose prompt TeV emission is strong. Since almost all the long 
GRBs accompany afterglows, the possible candidates of such bursts 
are likely to be a fraction of short GRBs that do
not show conventional X-ray afterglows (only show a steep decay phase
as the tail of prompt emission spectrum). In fact, $\sim 1/3$ of short GRBs
(e.g., GRB 050906, 051210, 070209, 070810B and 080121) are such ``naked" bursts
maybe due to the low density of the circumburst medium (e.g., La Parola et al. 2006). 
Since these bursts are spectrally hard and less energetic (than their long brethren), 
they may have prompt emission extending to the TeV range (e.g., Gupta \& Zhang 2007). 
These bursts could therefore be the best targets to detect the pair echoes or to use
non-detections to constrain the IGMF.

In Figs. 9 and 10, we show the resulting spectra and light 
curves of the prompt-induced pair echo from a nearby, rather 
energetic short GRB. The parameters for the 
primary prompt emission are taken as the following: 
$\mathcal{E}_{\gamma, >0.1~{\rm TeV}}={10}^{51.5}$ ergs, $\beta=2.2$ and 
$E_{\gamma}^{\rm cut}={10}^{0.5}$ TeV. The duration is set to $\Delta T^{\prime}
=1~{\rm s}$. For naked GRBs, we expect that the primary emission 
decays according to the curvature effect, which typically drops as 
$F_{\rm pri} \propto t^{-3}$. For instance, when $E_{\gamma} F_{\gamma} 
\sim {10}^{-2}~{\rm GeV}~{\rm cm}^{-2}~{\rm s}^{-1}$ during 
the burst, we have $E_{\gamma} F_{\gamma} 
< {10}^{-8}~{\rm GeV}~{\rm cm}^{-2}~{\rm s}^{-1}$ at $t > 100$ s.  
Hence, we omit the afterglow spectra/light curves in Figs. 9 and 10. 
As is seen in Fig. 10, the IGMF of $B_{\rm IG} \lambda_{\rm coh}^{1/2} 
\sim {10}^{-22}~{\rm G}~{\rm Mpc}^{1/2}$ leads to the detectable 
flux at $t \sim {10}^4~{\rm s}$, which should be observed 
as extended high-energy emission from short GRBs. 
Note that, when $B_{\rm IG} \sim 0$ G, the pair echo duration 
is determined by the angular spreading time, 
$300~{\rm s}~{(n_{\rm CIB}^{\prime}/0.1~{\rm cm}^{-3})}^{-1}$. 
Therefore, it may typically be difficult for pair echoes to explain   
GeV emission whose time scale is shorter (e.g., GRB 081024B 
and see also discussions in Zou, Fan, \& Piran 2008), 
but they may also generate the high-energy extended emission.

For non-detections, one may obtain a constraint as
\begin{equation}
B_{\rm IG} {\rm min[\lambda_{\rm coh}^{1/2},\lambda_{\rm IC}^{1/2}]} > {10}^{-21.5}~{\rm G}~{\rm Mpc}^{1/2}, 
\end{equation}
for our optimistic prompt parameters. 
We need to observe primary TeV emission for this purpose, but 
it is more difficult to make follow-up observations 
for short GRBs with MAGIC and VERITAS, compared to long GRBs.
Note that significant and non-tentative TeV signals have not 
been observed so far for both of the long and short GRBs 
\cite{Abd+07,Alb+07}. This may be because a part of GRBs 
can be TeV emitters due to the small optical thickness for 
pair creation and TeV photons from distant sources are 
significantly attenuated by the CIB.   

\section{Summary and Discussion}
In this paper, we have calculated the time-dependent 
spectra of the secondary pair echoes from the GRB prompt 
and afterglow TeV emission components that are attenuated by the CIB, 
applying a recently developed formalism to properly describe
the temporal evolution of the pair echoes.
We have compared the flux of the pair echoes to that of 
the afterglow, taking into account 
up-scattering of the CIB photons.
In particular, we have demonstrated (1) that afterglow-induced 
pair echoes can be important after the jet break for long
GRBs with a canonical afterglow; and 
(2) that prompt-induced pair echoes may also outshine the 
afterglow emission, if the prompt TeV emission is intense,
typically with $\mathcal{E}_{\gamma, >0.1~{\rm TeV}} > 
Y {(1+Y)}^{-1} \epsilon_e \mathcal{E}_k$. 

Weak but non-zero IGMFs can be crucial for detectability, 
since they make the duration of the pair echo emission much 
longer than the time scale of primary emission (see 
Figs. 2, 4, 6, and 8). Although the detectability itself 
also depends on both of the spectral evolution of the primary 
emission and detector sensitivities, such non-zero IGMFs can 
make it easier to detect secondary photons at late times 
when the pair echo emission remains shallow compared to 
the afterglow emission.  
Concerning with the detection of pair echo signals, 
``naked" (short) GRBs without a significant afterglow 
emission could be more promising. The pair echo 
should be observed as extended emission with 
the time scale of $t > 30-300$ s.
The observational prospects of 
such pair echoes are quite interesting
for the recently launched \textit{Fermi}. Successful
detections may be possible for nearby, bright 
events, and would open a new window to study the
poorly unknown IGMF. Even in the case of non-detections,
lower limits on the IGMF of $B_{\rm IG} \cdot 
{\rm min[\lambda_{\rm coh}^{1/2}, \lambda_{\rm IC}^{1/2}]} 
\sim {10}^{-20}-{10}^{-21}~{\rm G}~{\rm Mpc}^{1/2}$ 
may be obtained.

The main caveat in hunting afterglow-induced pair echoes 
and pair echoes from short GRBs is that nearby bright 
GRBs do not seem frequent. Although there is large uncertainty on the 
nearby burst rate, the rate of bursts occurring within $z \sim 0.3$ is 
estimated as $\sim$ a few events per year \citep[e.g.,][]{GPW05,GP06,LZVD07}. 
The actual detection rate also depends on several 
factors such as the detector sensitivity and field of 
view (e.g., $\sim 2.4$ sr for the \textit{Fermi}/LAT detector), so 
that only a fraction of them would be detected.
If all the bursts are ideal TeV emitters, we can expect 
pair echoes for these bursts in the near future. 
However, it is unlikely that all the bursts are bright TeV emitters 
(and it seems more plausible for prompt emission due to significant 
attenuation by the pair creation). 
Although it is currently impossible to predict how many bursts can be 
bright TeV emitters in both the prompt and afterglow phases,  
the expected detection rate for $z < 0.3$ bursts would be ``at most" 
$\sim 1-2$ events per year. 
There may be further complications about nearby GRBs.  
Some of the nearby long bursts detected so far seem somewhat dimmer 
than classical GRBs occurring at $z > 1$, but their local rate may be higher 
than the estimated local rate of classical GRBs \citep[e.g.,][]{GD07,LZVD07}. 
Hence, we may have more nearby bursts that can be detected 
in the keV-MeV band by detectors with better sensitivities 
(e.g., \textit{EXIST}). But, since the typical luminosity of such 
low luminosity bursts seems small, it is not so easy to see pair 
echoes from them. In addition, energetic short GRBs assumed 
in Figs. 9 and 10 would also be rare, whose radiation 
energy is larger than the typical one 
($\mathcal{E}_{\gamma}^{\rm iso} \sim {10}^{50-51}$ ergs).  
Nevertheless, possible detections of pair echoes would bring us a big impact
in understanding GRB physics and IGMF, 
even though the bright TeV GRBs that can lead to such detections are rare.
The current on-orbit \textit{Fermi} satellite 
is suitable for such a purpose. MAGIC and VERITAS can also provide valuable
data via follow-up observations, since the pair echo 
emission can last for a long duration of time. 
In the near future, some constraints on 
the models may be achieved even for non-detections.

We must also beware of the uncertainties in the
intrinsic primary spectra since the pair echo flux
depends on the amount of TeV photons. 
As for afterglow emission, we only consider the 
conventional forward shock model with energy injection. 
Although other parameter sets or 
other models such as the varying $\epsilon_e$ model 
can be considered, we expect that the
qualitative features of the pair echoes themselves
will not be changed significantly, as long as 
the light curve of high-energy emission 
is similar to that of X-rays and 
the amount of TeV photons is not too different 
from that invoked in our case.  
As for the prompt emission, possible uncertainties may come from 
the intrinsic emission properties such as $E_{\gamma}^{\rm cut}$, 
as discussed in Murase et al. (2007).

The contamination by other high-energy emission components 
might complicate the picture further.
There are many possibilities of 
high-energy gamma ray emission during the afterglow phase 
\citep[see, e.g.,][and references therein]{Zha07,FP08}.
For example, high-energy emissions associated with X-ray flares 
are expected at $\sim$ GeV energies. GeV photons 
can be produced by both of the leptonic mechanisms 
\citep[e.g.,][]{Wei+06,Wan+06,YD08} and the hadronic mechanisms 
\cite{MN06}. In addition, the reverse shock electrons can also 
provide high-energy photons during the early afterglow 
phase. Nonetheless, it is in principle possible to 
distinguish the pair echo emission from other possibilities, 
given an ideal broad-band (optical, X-ray, MeV and GeV)
observational campaign.

\section*{Acknowledgments}
KM and KT are supported by a Grant-in-Aid for the JSPS fellowship.
BZ acknowledges NASA NNG05GB67G, NNX08AN24G, and NNX08AE57A for 
support.
SN is supported in part by  Grants-in-Aid for Scientific Research
from the Ministry of E.C.S.S.T. (MEXT) of Japan, Nos. 19104006, 
19740139, 19047004. 
The numerical calculations were carried 
out on the Altix3700 BX2 at the YITP in Kyoto University.



\label{lastpage}


\begin{thebibliography}{99}
\bibitem[Abdo et al. 2007]{Abd+07}
Abdo, A.-A., et al. 2007, ApJ, 666, 361
\bibitem[Albert et al. 2007]{Alb+07}
Albert, J. et al. 2007, ApJ, 667, 358
\bibitem[Albert et al. 2008]{Alb+08} 
Albert, J. et al. 2008, Science, 320, 1752
\bibitem[Ando 2004]{And04}Ando, S. 2004, MNRAS, 354, 414
\bibitem[Asano \& Inoue 2007]{AI07} 
Asano, K, \& Inoue, S. 2007, ApJ, 671, 645

\bibitem[Beloborodov 2005]{Bel05}
Beloborodov, A.-M. 2005, ApJ, 618, L13

\bibitem[Carmona et al. 2007]{Car+07}
Carmona, E. et al. 2007, ArXiv e-prints, arXiv:0709.2959

\bibitem[Dai \& Lu 2002]{DL02}
Dai, Z.-G., \& Lu, T. 2002, ApJ, 580, 1013
\bibitem[Dai et al. 2002]{Dai+02}
Dai, Z.-G., Zhang, B., Gou, L.-J., Meszaros, P., \& Waxman, E. 2002, ApJ, 580, L7

\bibitem[Eichler \& Granot 2006]{EG06}
Eichler, D., \& Granot, J. 2006, ApJ 641, L5 

\bibitem[Fan \& Piran 2008]{FP08}
Fan, Y.-Z., \& Piran, T. 2008, Frontiers of Physics in China, 3, 306
\bibitem[Fan et al. 2008]{Fan+08} 
Fan, Y.-Z., Piran, T., Narayan, R., \& Wei, D.-M. 2008, MNRAS, 384, 1483
\bibitem[Furlanetto \& Loeb 2001]{FL01}
Furlanetto, S.-R., \& Loeb, A. 2001, ApJ, 556, 619

\bibitem[Genet et al. 2007]{GDM07}
Genet, F., Daigne, F., Mochkovitch, R. 2007, MNRAS, 381, 732
\bibitem[Ghisellini et al. 2007]{Ghi+07}
Ghisellini, G. et al. 2007, ApJ, 658, L75
\bibitem[Gnedin et al. 2000]{GFZ00}
Gnedin, N.-Y., Ferrara, A., \& Zweibel, E.-G. 2000, ApJ, 539, 505
\bibitem[Gou \& M\'esz\'aros 2007]{GM07}
Gou, L.-J., \& M\'esz\'aros, P. 2007, ApJ, 668, 392
\bibitem[Granot et al. 2008]{Gra+08}
Granot, J., Cohen-Tanugi, J., do Couto e Silva, E. 2008, ApJ, 677, 92  
\bibitem[Guetta \& Granot 2003]{GG03}
Guetta, D., \& Granot, J. 2003, ApJ, 585, 885
\bibitem[Guetta et al. 2005]{GPW05}
Guetta, D., Piran, T., \& Waxman, E. 2005, ApJ, 619, 412
\bibitem[Guetta \& Piran 2006]{GP06}
Guetta, D., \& Piran, T. 2006, A\&A, 453, 823
\bibitem[Guetta \& Della Valle 2007]{GD07}
Guetta, D., \& Della Valle, M. 2007, ApJ, 657, L73 
\bibitem[Gupta \& Zhang 2007]{GZ07}
Gupta, N., \& Zhang, B. 2007, MNRAS, 380, 78
\bibitem[Gupta \& Zhang 2008]{GZ08}
Gupta, N., \& Zhang, B. 2008, MNRAS, 384, L11

\bibitem[Hurley et al. 1994]{Hur+94}
Hurley, K. et al. 1994, Nat, 372, 652

\bibitem[\protect\citeauthoryear{Ichiki et al.}{2008}]{IIT08}
Ichiki, K., Inoue, S., \& Takahashi, K. 2008, ApJ, 682, 127
\bibitem[Ichiki et al. 2006]{Ich+06}
Ichiki, K., Takahashi, K., Ohno, H., Hanayama, H., \& Sugiyama, N. 2006,
Science, 311, 827

\bibitem[Kneiske et al. 2002]{Kne+02}
Kneiske, T.-M., Mannheim, K., \& Hartmann, D.-H. 2002, A\&A, 386, 1
\bibitem[Kneiske et al. 2004]{Kne+04}
Kneiske, T.-M. et al. 2004, A\&A, 413, 807
\bibitem[Kronberg 1994]{Kro94}
Kronberg, P.-P. 1994, Rep. Prog. Phys., 57, 325

\bibitem[La Parola et al. 2006]{LaP+06}
La Parola, V. et al. 2006, A\&A, 454, 753

\bibitem[Liang et al. 2008]{Liang08}
Liang, E.-W. et al. 2008, ApJ, 675, 528
\bibitem[Liang et al. 2007a]{LZZ07}
Liang, E.-W., Zhang, B.-B., Zhang, B. 2007a, ApJ, 670, 565
\bibitem[Liang et al. 2007b]{LZVD07}
Liang, E.-W., Zhang, B., Virgili, F., Dai, Z.-G. 2007b, ApJ, 662, 1111
\bibitem[Lithwick \& Sari 2001]{LS01}
Lithwick, Y., \& Sari, R. 2001, ApJ, 555, 540

\bibitem[M\'esz\'aros 2006]{Mes06}
M\'esz\'aros, P. 2006, Rep. Prog. Phys., 69, 2259

\bibitem[Murase \& Nagataki 2006]{MN06} 
Murase, K, \& Nagataki, S. 2006, Phys. Rev. Lett., 97, 051101
\bibitem[Murase \& Ioka 2008]{MI08}
Murase, K., \& Ioka, K. 2008, ApJ, 676, 1123 
\bibitem[Murase et al. 2007]{MAN07} 
Murase, K., Asano, K, \& Nagataki, S. 2007, ApJ, 671, 1886
\bibitem[Murase et al. 2008a]{Mur+08a}
Murase, K., Ioka, K., Nagataki, S., \& Nakamura, T. 2008, Phys. Rev. D, 78, 023005 
\bibitem[Murase et al. 2008b]{Mur+08}
Murase, K., Takahashi, K., Inoue, S., Ichiki, K., \& Nagataki, S. 2008, ApJ, 686, L67 

\bibitem[Nousek et al. 2006]{Nou+06}
Nousek, J.-A. et al. 2006, ApJ, 642, 389

\bibitem[O'Brien et al. 2006]{OBr+06}
O'Brien, P.-T. et al. 2006, ApJ, 647, 1213

\bibitem[Panaitescu 2007]{Pan07}
Panaitescu, A. 2007, ApJ, 379, 331

\bibitem[Plaga 1995]{Pla95}
Plaga, R. 1995, Nature, 374, 30
\bibitem[Razzaque et al. 2004]{RMZ04}
Razzaque, S., M\'esz\'aros, P., \& Zhang, B. 2004, ApJ, 613, 1072
\bibitem[Rhoads 1999]{Rhoads99}
Rhoads, J. E. 1999, ApJ, 525, 737

\bibitem[Sari \& Esin 2001]{SE01}
Sari, R., \& Esin, A.-A., 2001, ApJ, 548, 787 

\bibitem[Sari et al. 1999]{SPH99}
Sari, R., Piran, T., Halpern, J. 1999, ApJ, 519, L17

\bibitem[Stecker \& Scully 2008]{SS08}
Stecker, F.-W., \& Scully, S.-T. 2008, arXiv:0807.4880

\bibitem[Takahashi et al. 2005]{Tak+05}
Takahashi, K., Ichiki, K., Ohno, H., \& Hanayama, H. 2005, Phys. Rev. Lett.,
95, 121301
\bibitem[Takahashi et al. 2008]{Tak+08} 
Takahashi, K, Murase, K., Ichiki, K., Inoue, S., \& Nagataki, S. 2008, ApJ, 687, L5

\bibitem[Turner \& Widrow 1988]{TW88}
Turner, M.-S., \& Widrow, L.-M. 1988, Phys. Rev. D, 37, 2743

\bibitem[Uhm \& Beloborodov 2007]{UB07}
Uhm, Z.-L., \& Beloborodov, A.-M. 2007, 665, L93 

\bibitem[Wang et al. 2004]{Wan+04}
Wang, X.-Y., Cheng, K.-S., Dai, Z.-G., \& Lu, T. 2004, ApJ, 604, 306
\bibitem[Wang et al. 2006]{Wan+06}
Wang, X.-Y., Li, Z., \& M\'esz\'aros, P. 2006, ApJ, 641, L89
\bibitem[Wei et al. 2006]{Wei+06}
Wei D.-M., Yan T., \& Fan Y.-Z. 2006, ApJ, 636, L69
\bibitem[Wei \& Fan]{WF07}
Wei, D.-M., \& Fan, Y.-Z. 2007, ChJAA, 7, 509
\bibitem[Widrow 2002]{Wid02}
Widrow, L.-M. 2002, Rev. Mod. Phys. , 74, 775

\bibitem[Yamazaki 2009]{Yam09}
Yamazaki, R. 2009, ApJ, 690, L118
\bibitem[Yu \& Dai 2008]{YD08}
Yu, Y.-W., \& Dai, Z.-G. 2008, ArXiv e-prints, arXiv:0811.1068

\bibitem[Zhang et al. 2006]{Zha+06}
Zhang, B., et al. 2006, ApJ, 642, 354
\bibitem[Zhang 2007]{Zha07}
Zhang, B. 2007, ChJAA, 7, 1
\bibitem[Zhang \& M\'esz\'aros 2001]{ZM01}
Zhang, B., \&  M\'esz\'aros, P. 2001, ApJ, 559, 110
\bibitem[Zou, Fan, \& Piran 2008]{ZFP08}
Zou, Y.-C., Fan, Y.-Z., \& Piran, T. 2008, ArXiv e-prints, arXiv:0811.2997
\end{thebibliography}
\end{document}